\documentclass[a4paper,titlepage,fontsize=11pt,DIV=12,headings=normal]{scrartcl}
\usepackage[affil-it]{authblk}
\usepackage{moreverb}

\usepackage[colorlinks,bookmarksopen,bookmarksnumbered,citecolor=red,urlcolor=red]{hyperref}

\newcommand\BibTeX{{\rmfamily B\kern-.05em \textsc{i\kern-.025em b}\kern-.08em
T\kern-.1667em\lower.7ex\hbox{E}\kern-.125emX}}

\usepackage{url}
\usepackage{epsfig}
\usepackage{amssymb}
\usepackage{amsmath}
\usepackage{amsfonts}

\usepackage{xargs}                      
\usepackage[pdftex,dvipsnames]{xcolor}  

\usepackage{adjustbox}
\usepackage{array}
\usepackage{booktabs}
\usepackage{multirow}

\newcolumntype{R}[2]{%
    >{\adjustbox{angle=#1,lap=\width-(#2)}\bgroup}%
    l%
    <{\egroup}%
}
\newcommand*\rot{\multicolumn{1}{R{90}{1em}}}

\begin{document}



\huge
\begin{centering}
Micro protocol engineering for unstructured carriers:\\On the embedding of steganographic control protocols into audio transmissions

~

\normalsize

\large
Matthias Naumann$^1$, Steffen Wendzel$^2$, Wojciech Mazurczyk$^3$, J{\"o}rg Keller$^1$

~

\normalsize
$^1$ University of Hagen, Germany\\
$^2$ Fraunhofer FKIE, Germany\\
$^3 $Warsaw University of Technology, Poland\\
~
\end{centering}


\normalsize

\textbf{Abstract.}
Network steganography conceals the transfer of sensitive information within unobtrusive data in computer networks. So-called micro protocols are communication protocols placed within the payload of a network steganographic transfer. They enrich this transfer with features such as reliability, dynamic overlay routing, or performance optimization --- just to mention a few. We present different design approaches for the embedding of hidden channels with micro protocols in digitized audio signals under consideration of different requirements. On the basis of experimental results, our design approaches are compared, and introduced into a protocol engineering approach for micro protocols.

~

\textbf{Keywords.} network steganography; audio steganography; micro protocols; control protocols; covert channels

~


\section{Introduction}

Network steganography is the art and science of transferring secret information over a network while the hidden transfer itself is concealed. Network steganography is applied when the encryption of secret data is not sufficient \cite{ZanderSurvey}. For instance, journalists can use network steganography to keep the transfer of illicit information hidden when facing Internet censorship. In addition to the transfer of hidden payload, so-called micro protocols (or control protocols) can be embedded into a network steganographic communication. Micro protocols enrich the feature-set of the steganographic channel in various aspects \cite{ANTEArticle}, such as
\begin{itemize}
 \item adapting the steganographic channel's configuration automatically to bypass new Internet censorship infrastructure,
 \item integrate functionality for reliability and segmentation of hidden data, or
 \item provide dynamic overlay routing and optimization for the stealthiness of a hidden transmission.
\end{itemize}

Being steganographic payload itself, a poor micro protocol design and placement can lead to a higher detectability of a steganographic communication. This problem led to the development of protocol engineering methods for micro protocols by Wendzel, Keller and Backs in \cite{ANTEArticle,BacksEtAl,CMS12}. These protocol engineering methods either minimize the size of the micro protocol or the placement within a \emph{structured} carrier (machine readable, such as protocol header fields) but not the placement within an \emph{unstructured} carrier (understandable by humans, such as audio content in audio or video transmissions).

Our work presents the first design approaches, evaluation and protocol engineering approach for micro protocols to be placed in unstructured carriers. We demonstrate our approach on the basis of digital audio transmissions.

Micro protocols comprise their own protocol headers. These headers can be designed in two ways, either in form of a

\begin{enumerate}
\item \emph{static header}, i.e., the whole header is present in one chunk, or in form of a
\item \emph{dynamic header}, i.e. the header can be split into separate parts which are included on demand and spread over the carrier.
\end{enumerate}

For our micro protocol design, we consider both static and dynamic headers. Moreover, our work examines how micro protocols must be designed to provide a high level of covertness (stealthiness) in unstructured carriers. For the application of micro protocols to hidden data within audio streams criteria like covertness, capacity and robustness of the transmitted data are considered with respect to the specific conditions of a transmission of audio data and the human auditive perception.

The reminder of this paper is structured as follows. Section~\ref{Sect:Fundamentals} explains fundamentals of related topics and discusses related work on micro protocols. In Section~\ref{Sect:ReqAnal}, we perform the requirements analysis for our micro protocol engineering approach, followed by Section~\ref{Sect:Implementation} in which we discuss the implementation of the approach. Section~\ref{Sect:Evaluation} evaluates our work on the basis of experiments. On the basis of this evaluation, Section~\ref{Sect:MPEngAppr} presents our protocol engineering approach. Section~\ref{Sect:Concl} concludes and discusses future work.

\section{Fundamentals}\label{Sect:Fundamentals}

In this section, we first introduce basics of network steganography and related topics which are of importance for this paper. We finish with an introduction to micro protocols and a discussion of related work.

\subsection{Steganography and Its Detection}

The goals of a steganographic transfer can vary from the theft of confidential data, e.g., within industrial espionage, over the hidden communication of command and control data in botnets, to the transmission of illicit information by journalists \cite{ZanderSurvey}. Steganography dates back to ancient Greece. Early approaches for steganography included the hiding of information using e.g. an invisible ink. In the late 20th century, digital media steganography arose, which focuses on hiding information in digital audio, image, and video files \cite{Zielinska2014}.

In \emph{digital audio steganography}, confidential data is embedded into digital sound. Therefore, bit sequences of the digital audio files are slightly changed to encode hidden data. This embedding of data into audio files is especially challenging since the human auditive perception is highly sensitive, in particular even more sensitive than the human visual system \cite{n5:perception}. Research focuses on the development of suitable technologies to prevent the perception of the steganographic data in audio files by both, humans and machines. In \cite{n6:KatzenbeisserBook}, some of the most widely used algorithms for audio steganography are discussed, which are LSB (\emph{Least Significant Bit}) Coding, Phase Coding, Parity Coding, Spread Spectrum and Echo Data Hiding.
Transferring hidden data in audio data shares some aspects with audio watermarking (cf.\ e.g.\ \cite{KM2003}), such as close to no influence on perception by humans, and survival of operations like filtering. The difference is that audio watermarks are not necessarily hidden data, and that they are contempt if enough data from the watermark remains to identify the watermark, while enough of the hidden data (including error-correcting code) must remain in order to completely and correctly decode the payload. 

\emph{Network steganography} stealthily transfers arbitrary data over a network. Therefore, either structured or unstructured carriers can be used. \emph{Structured} data are those interpretable by machines, having a pre-defined structure, such as a network protocol header, while \emph{unstructured} data is interpretable by a human, e.g. the content of an audio stream, an image file or a natural language \cite{FiskEtAl}. For instance, if data is hidden within the IPv4 TTL field, it is hidden in structured data because protocol headers always have a pre-defined, machine-interpretable structure. The embedding of hidden data in inaudible parts of audio streams, on the other hand, is a form of network steganography in unstructured data.


From the audio steganography perspective, many methods (both for data hiding as well as for their detection) have been designed and investigated e.g. for services like VoIP (Voice over IP). As in this paper we investigate micro protocols for VoIP-like service that utilize the LSB algorithm as an underlying steganographic method thus below we review related work that describes LSB-based solutions and their detection.
Currently, a lot of research effort is still devoted for improving LSB steganography for real-time services like VoIP. 	The first VoIP steganographic method to utilize the digital voice signal as a hidden data carrier was proposed by Aoki in \cite{Aoki2003}. The LSB steganography was utilized to provide a PLC (Packet Loss Concealment) method for G.711-based VoIP. 
	Dittmann et al.\ in 2005 presented the first VoIP steganography implementation prototype that also used the LSB method \cite{Dittmann2005}. 
	Wu and Yang described the scheme of adaptive LSB \cite{Wu2006}, which for G.711-based speech calculates energy statistics to estimate the number of least significant bits to be utilized as a hidden data carrier in each voice sample. The results proved that this approach performs better than simple LSB and offers a higher steganographic bandwidth (about 20 kbit/s) while introducing less degradation of the voice quality.
	Wang and Wu also suggested using the least significant bits of voice samples to carry secret communication \cite{Wang2007}, but in their solution the bits of the steganogram were coded using a low rate voice codec, like Speex. Their prototype implementation is characterized by a small processing delay of about 0.257 ms.
	Takahashi and Lee presented a proof of concept LSB-based implementation, Voice over VoIP (Vo2IP), which is able to establish a hidden communication by embedding 8 kbit/s G.729-based compressed voice data into the regular PCM (Pulse Code Modulation)-based voice traffic \cite{Takahashi2007}.
	In 2008 Liu et al.\ found that the LSBs of each speech frame for G.729 can be replaced with secret data bits \cite{Liu2008}. The experimental results indicate that the method is perceptually transparent while the steganographic bandwidth is relatively high (about 200 bit/s).
	Tian et al.\ proposed the use of an LSB steganography-based system that employs a well-balanced and simple encryption of secret data \cite{Tian2008}. This system was evaluated for VoIP with G.729a speech coding using a proof of concept tool named \emph{StegTalk}. The experimental results showed that the achievable steganographic bandwidth is in the range 0.8–2.6 kbit/s and has a negligible effect on speech quality. Moreover, it met the real-time requirements of the VoIP service.
	A real-time steganography system for VoIP was described by Tian et al.\ in \cite{Tian2009}. The main novelty of the proposed solution is not in the steganographic method used (LSB), but in utilizing M-sequence encryption techniques to eliminate the correlation amongst secret messages to increase the resistance against statistical steganalysis. Moreover, protocol steganography (usage of free/unused fields in protocols headers) is applied to provide a novel synchronization mechanism together with an RSA-based key agreement that ensures accurate restitution of the secret messages on the receiver side. This system was experimentally evaluated for 0.8 and 2.6 kbit/s steganographic bandwidth, which obtained a 0.3 and 1 quality drop in the MOS (Mean Opinion Score) scale (which is typically used for expressing the quality of VoIP calls), respectively and the total embedding latency increased by about 4.7 ms when 1 MB of steganogram is transmitted.
	In 2009 Xu and Yang proposed an LSB-based method dedicated to voice transmission using the G.723.1 codec in 5.3 kbit/s mode \cite{Xu2009}. They identified five least significant bits of the LSP VQ (Line Spectrum Pair Vector Quantization) indices and used them to transmit hidden data; the method provided a steganographic bandwidth of 133.3 bit/s. 
	An insightful overview of the general techniques that can be applied to VoIP steganography methods to make their detection even more difficult was introduced by Tian et al.\ in \cite{Tian2011}. Additionally, they proposed three new encoding strategies based on digital logic. All techniques were evaluated for LSB-based steganography and proved to be effective.
	Another adaptive LSB-based steganography approach named AVIS (\emph{Adaptive VoIP Steganography}) was proposed by Xu et al.\ \cite{Xu2011}. AVIS has two components: VAMI (\emph{Value-based Multiple Insertion}), which is responsible for dynamically selecting multiple bits based on the VoIP vector value and VADDI (\emph{Voice Activity Detection Dynamic Insertion}), which dynamically changes the embedding intervals to make detection harder. The approach was implemented for G.711-based VoIP and the results prove that it is less detectable than a classic LSB method whilst achieving a steganographic bandwidth of about 114 B/s, introducing acceptable delay and degrading the voice from 0.1 to 0.4 on MOS scale.
	Finally, Liu et al.\ adopted least-significant-digits rather than LSBs to hide secret data \cite{Liu2012}. This approach can increase around 30\% of steganographic bandwidth while introducing lower steganographic cost than classic LSB method.

From the covert communication detection perspective statistical steganalysis for LSB-based VoIP steganography was first proposed by Dittmann et al.\ \cite{Dittmann2005}. They proved that it was possible to detect hidden communication with almost a 99\% success rate under the assumption that there are no packet losses and the steganogram is unencrypted/uncompressed.
	Takahasi and Lee described a detection method based on calculating the distances between each audio signal and its de-noised residual by using different audio quality metrics \cite{Takahashi2007}. Then, a SVM classifier is utilized for detection of the existence of hidden data. This scheme was tested on LSB, DSSS, FHSSS and Echo hiding methods and the results obtained show that for the first three algorithms the detection rate was about 94\% and for the last it was about 73\%.
	A Mel-Cepstrum based detection, known from speaker and speech recognition, was introduced by Kraetzer and Dittmann \cite{Kraetzer2007} for the purpose of VoIP audio steganalysis. Under the assumption that a steganographic message is not permanently embedded from the start to the end of the conversation, the authors demonstrated that the detection of an LSB-based steganography is efficient with a success rate of 100\%.
	Steganalysis of LSB steganography based on a sliding window mechanism and an improved variant of the previously known Regular Singular (RS) algorithm was proposed by Huang et al.\ \cite{Huang2011c}. Their approach provides a 64\% decrease in the detection time over the classic RS, which makes it suitable for VoIP. Moreover, experimental results prove that this solution is able to detect up to five simultaneous VoIP covert channels with a 100\% success rate. Huang et al.\ also introduced the steganalysis method for compressed VoIP speech that is based on second order statistics \cite{Huang2011d}. In order to estimate the length of the hidden message, the authors proposed to embed hidden data into a sampled speech at a fixed embedding rate, followed by embedding other information at a different level of data embedding. Experimental results showed that this solution not only allows the detection of hidden data embedded in a compressed VoIP call, but also to accurately estimate its size. 

Hanspach and Goetz present another acoustic covert channel that uses near-ultrasonic sound from a laptop speaker to another laptop's microphone to transfer data \cite{HG2013}. In their work, the acoustic signals themselves shall be unnoticed, while our research addresses hiding of information in digitized acoustic signals, which themselves are not hidden.

In general, steganographic traffic with micro protocols and without micro protocols can both be detected the same way. However, there is work on micro protocol-specific countermeasures that was published by Kaur et al. \cite{KaurEtAl:MPCountermeasures}. The authors demonstrate different active and passive attacks on two micro protocols, including attacks on dynamic overlay routing and sniffing attacks. Especially, if designed naively micro protocols can impact covertness negatively.

\subsection{Micro Protocols and Related Work}

\emph{Micro protocols} (also \emph{covert channel control protocols} or \emph{steganographic control protocols}) are protocols featuring headers, which are placed into a steganographic carrier. In other words, they are part of the hidden data transferred in a steganographic transmission and share the available space with the hidden payload of the transmission.

A number of micro protocols exist, which are surveyed and compared in \cite{ANTEArticle}. The provided features of micro protocols differ highly, for instance, \cite{CFTP} presents a micro protocol with application layer functionality that provides a file transfer service, while \cite{BacksEtAl} presents a dynamic routing protocol for steganographic overlay networks. Other common features are to provide reliability for the transfer of steganographic payload or to alternate between different hiding methods on demand to overcome filtering technology in censorship environments.

As micro protocols share the usually limited space within a steganographic carrier with the actual payload, two requirements arise \cite{ANTEArticle}:

\begin{enumerate}
 \item The micro protocol header must be as small as possible so that more payload can be transferred per packet as the covertness of a connection increases when less hidden data must be transferred.
 \item The micro protocol header must be embedded into the steganographic carrier in a way that its placement is always conforming to the carrier. Otherwise, the micro protocol may cause unusual states in the carrier (e.g. unusual flag combinations of headers), resulting in lower covertness.
\end{enumerate}

Only two micro protocol engineering approaches exist and both focus on different goals. Backs et al.\ address the first requirement of a minimized header size in \cite{BacksEtAl}. For this reason, they introduce a dynamic micro protocol header based on the concept of \emph{status updates}. Each header component of the micro protocol is only embedded on demand to reduce the overall size. If a header component indicating a `state', such as the destination of the attached payload, does not change, the header component indicating the state is not required to be sent again until its state changes. In other words, a status update is performed only on demand. Moreover, the micro protocol header can be fragmented into an arbitrary number of pieces.

The second approach, which consists of six steps, was presented by Wendzel and Keller in \cite{CMS12} and addresses the second requirement: it maximizes the covertness of the micro protocol's placement and behavior. Their approach focuses solely on structured carriers, namely network protocol headers. The ocurrence rates of bits in the micro protocol and bits of the carrier are mapped in a way that the placement of the micro protocol causes as few anomalies as possible. In \cite{ANTEArticle}, the approach was optimized in order to reflect bit mappings based on protocol states. For instance, bit mappings of a micro protocol can depend on the state of a TCP connection if embedded into TCP.

Moreover, the carrier and the micro protocol can each be described in form of a formal grammar \cite{CMS12}. Using a language inclusion test, Wendzel and Keller verify whether the micro protocol's design is conforming to the design of the carrier. For example, some bits of carrier headers can only occur at specific phases of a connection and not at the same time as other bits. If the micro protocol violates such rules, the resulting anormal carrier behavior would rise suspicion.

The only micro protocol that partly embeds into unstructured network data was presented by Mazurczyk and Kotulski in \cite{HybridPaper} but embeds major information into a structured part of the RTP header and provides no micro protocol engineering approach.

Thus, while Backs et al.\ reduce the size of the micro protocol, Wendzel and Keller optimize the design and embedding of the protocol for structured carriers. No approach is available for the optimal design and placement of micro protocols into unstructured carriers, which is subject of this work.

\section{Requirements Analysis}\label{Sect:ReqAnal}

In this section, we will first introduce the requirements for the embedding of a micro protocol into audio streams. We describe the implementation details in the following section. Based on our evaluation, we will conduct the actual micro protocol engineering approach for unstructured carriers. The implementation and evaluation are a necessity as no evaluation for the embedding of different micro protocol designs in unstructured carriers exists.

Basic requirements for the embedding of a micro protocol are the non-functional requirements of maximized covertness, robustness against changes, high channel capacity and a low overhead of the steganographic transmission.

We achieve a high covertness by a space-efficient header design: the fewer bits must be hidden, the smaller the influence on the utilized carrier. Therefore, our micro protocol engineering approach must result in a space-optimized protocol design. For this reason, we apply the concept of the aforementioned status updates. 

For the receiver, it must be possible to extract the hidden information of network packets at any time. Two different ways exist to embed a micro protocol into an unstructured carrier:

\begin{enumerate}
\item \emph{Fixed embedding:} the micro protocol header is placed at a fixed position within the audio data.
\item \emph{Dynamic embedding:} the first bits of the micro protocol header are placed at a fixed position within the audio data. These first bits indicate the placement of the following micro protocol header components within the same packet. In this scenario, the micro protocol header can be split into abitrary parts.
\end{enumerate}

In both cases, the receiver must know the position and length of (the first part of) the embedded micro protocol header. We decided to implement both approaches to enable their comparison. Moreover, the micro protocol receiver must be capable to detect whether a micro protocol is present in a received network packet, or not. For this reason, our approach requires the continuous sending of dummy data in every network packet to indicate the presence of a micro protocol.


We have chosen the simple LSB algorithm, which hides data in least significant bits, for the implementation of our proof of concept system. We used LSB as our work does not focus on the hiding algorithm of the covert data but on the design and placement of the micro protocol itself.
Hence, in a real implementation an algorithm would be used that is better suited against steganalysis but provides a similar quality, i.e.\ does not detoriate the audio quality notably.

\section{Implementation}\label{Sect:Implementation}

This section describes the micro protocol header design. The design is implemented in our proof of concept system and is evaluated in the following section.

The design of a micro protocol header can be performed in two ways. Either, the whole micro protocol header is transferred within a single network packet (refered to as a \emph{static} design). Or, the micro protocol is split over multiple packets were the earlier packets indicate that micro protocol header parts follow with the next packet (refered to as \emph{dynamic} design). We implement both design approaches in order to compare them and implemented them additionally using fixed embedding (see previous section). The combination of these design and embedding approaches allows a highly dynamic header structure and embedding. 

We decided to implement a status update-based micro protocol header consisting of \emph{attributes}. Each micro protocol header starts with an attribute called \emph{header type} (HT) that defines the structure of the remaining header.

\subsection{Design of a Static Micro Protocol Header}

Static headers comprise the advantage of a clear structure. On the other hand, static headers cannot be split over different network packets and they may comprise header components which are not required in every transferred packet. We defined the following four header types for a generic, static micro protocol header:

\begin{enumerate}
	\item REQ: a packet that initiates a request (incl. following data packets of a request),
	\item DAT: a data packet containing payload,
	\item RES: a response packet, containing no data as, in our model, responses only indicate whether a request was successful (ACK) or not (ERR), and
	\item DMY: a dummy type to indicate the presence of a micro protocol without content.
\end{enumerate}

The header type (HT) is followed by a number of other attributes as shown in Tab.~\ref{tab:StaticHdrFormat}, that also defines which attribute is included in each header type. The attributes are defined as follows:

\begin{itemize}
	\item NHO: an offset indicating the start of the next header attribute within the following packet,
	\item FMT: specification of the data format in a request (ASCII plaintext or binary data),\footnote{Responses do not contain payload (DAT) but the bi-directional communication allows every peer to transfer REQ packets containing payload.}
	\item CNT: counter for data packets (DAT header type),
	\item VER: version of the micro protocol used (only dynamic header),
	\item LEN: the length of attached data  (only dynamic header),
	\item CMD: specifies a command (`OK' or `RESEND') to indicate the need of a re-transmission, and
	\item DMY: indicating 9 bits of dummy data.
\end{itemize}

\begin{table*} \centering
    \begin{tabular}{llllcccc}
        \textbf{Attribute} & \textbf{Bits}~ &  ~\textbf{Comment}  & \textbf{Value(s)}~ & \rot{Request} & \rot{Data} & \rot{Response} & \rot{Dummy} \\
        \toprule
        \textbf{HT} header type 	& 2 	& \textbf{REQ} request 		& 00 		& ~$\bullet$~ 	& ~~~ 		& ~~~ 		& ~~~		 \\
        				& 	& \textbf{DAT} data		& 01		& 		& ~$\bullet$~	&		&		 \\
        				& 	& \textbf{RES} response		& 10		& 		& 		& ~$\bullet$~	&		 \\
        				& 	& \textbf{DMY} dummy		& 11		& 		& 		&		& ~$\bullet$~	 \\
        \hline
        \textbf{NHO} next header offset & 5	&				& 00000$\ldots$11111 & ~$\bullet$~ 	& ~$\bullet$~ 	& ~$\bullet$~ 	& ~$\bullet$~ 	 \\
        \hline
        \textbf{FMT} data format	& 1	& textual data			& 0		& ~$\bullet$~ 	& ~~~ 		& ~~~ 		& ~~~		 \\
        				& 	& binary data			& 1		& ~$\bullet$~ 	& ~~~ 		& ~~~		& ~~~		 \\
        \textbf{CNT} packet count	& 6	& ~~~			& 000000$\ldots$111111 & ~$\bullet$~ & ~~~	& ~~~ 		& ~~~~		 \\
        \textbf{VER} protocol version	& 2	& 				& 01-11		& ~$\bullet$~ 	& ~~~ 		& ~~~		& ~~~		 \\
        \hline
        \textbf{LEN} data length	& 8	& 				& 0000$\ldots$1111 & 	 	& ~$\bullet$~	& ~~~		& ~~~		 \\
        \hline
        \textbf{CMD} command		& 2	& OK				& 00 		& 	 	& 		& ~$\bullet$~	& ~~~		 \\
        				& 	& RESEND			& 01 		& 	 	& 		& ~$\bullet$~	& ~~~		 \\
	\hline
        \textbf{DMY} dummy data		& 9	& OK				& 9 random bits	& 	 	& 		& 		& ~$\bullet$~	 \\
        \bottomrule
    \end{tabular}
    \caption{Structure of our static micro protocol header}
    \label{tab:StaticHdrFormat}
\end{table*}

\begin{figure*}[t]
\centering
\epsfig{file=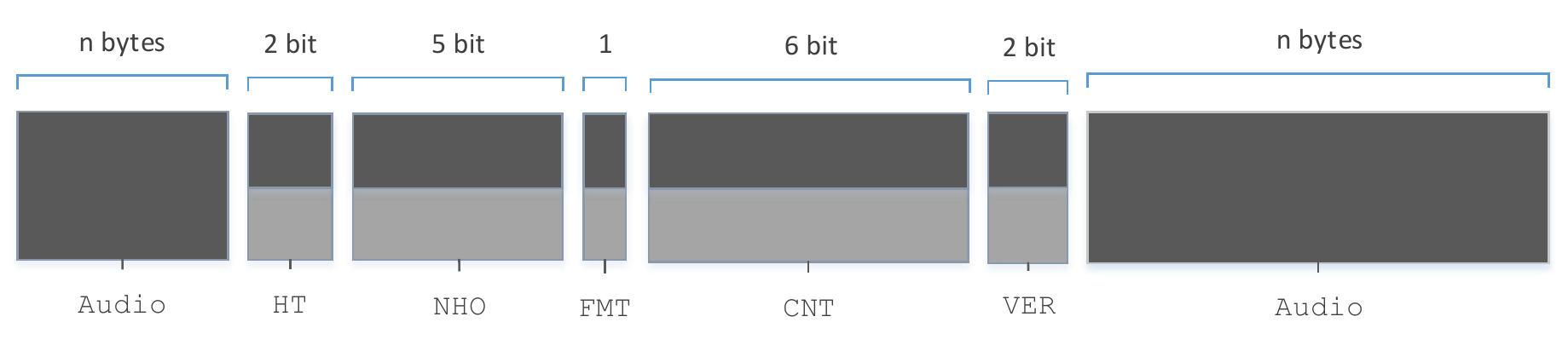, width=0.95\textwidth}
\caption{Example for a request using our static micro protocol header.}
\label{fig:StaticHdrExmpl}
\end{figure*}

Fig.~\ref{fig:StaticHdrExmpl} visualizes a sample packet for a static micro protocol header that is of the header type `request'. The whole header is embedded into the audio data transferred over the network. The header type starts at a fixed position (fixed embedding) and indicates a request (REQ), which is followed by the next header offset, the format bit, and the micro protocol's version number.

\subsection{Design of a Dynamic Micro Protocol Header}

Dynamic header designs allow to split headers into an abitrary number of data chunks. These data chunks can be embedded at different locations of an audio payload. While dynamic header designs result in higher implementation cost they also require the re-structuring of our previously introduced static header. We apply the same attribute types as for the static header but most of these attribute types are now header types (Tab.~\ref{tab:DynHdrFormat}).


We distinguish between primary and secondary header types. Primary header types indicate the overall meaning of a header (request, response or dummy data) while secondary header types depend on the primary header types. Secondary header types add additional attributes to the primary header type. In other words, secondary header types fullfil the same role as attributes in the \emph{static} header design but they can now be \emph{included on demand} and split over multiple locations within the audio carrier (by using their header type value to indicate their presence).

\begin{table*} \centering
    \begin{tabular}{lcllccc}
        \textbf{Attribute} & \textbf{Bits}~ &  ~\textbf{Description}  & \textbf{HT value}~ & \rot{Request} & \rot{Response} & \rot{Dummy} \\
        \toprule
        \textbf{HT} header type		& 3 	& \textbf{REQ} request 		& 000 		& ~$\bullet$~ 	& ~~~ 		& ~~~ 		 \\
        				& 	& \textbf{RES} response		& 001		& 		& ~$\bullet$~	&		 \\
        				& 	& \textbf{DMY} dummy		& 010		& 		& 		& ~$\bullet$~	 \\
        \cline{3-7}
        				&	& \textbf{DAT} data		& 011		& ~$\bullet$~	& 		&		 \\
        				& 	& \textbf{LEN} data length	& 100		& ~$\bullet$~	& 		&		 \\
        				& 	& \textbf{NHO} next header offset (optional) & 101 & ~$\bullet$~& 		& 		\\
        				& 	& \textbf{FMT} data format (optional)	& 110		& ~$\bullet$~	& 		&		 \\
        				& 	& \textbf{VER} protocol version (optional)	& 111		& ~$\bullet$~	& 		&		 \\
        \hline
        \textbf{VAL} value 		& 1$\ldots$n~ & allows to transfer payload	&  		& ~$\bullet$~ 	& ~$\bullet$~ 		& ~$\bullet$~ 		\\
        \bottomrule
    \end{tabular}
    \caption{Structure of our dynamic micro protocol header}
    \label{tab:DynHdrFormat}
\end{table*}

\begin{figure*}[t]
\centering
\epsfig{file=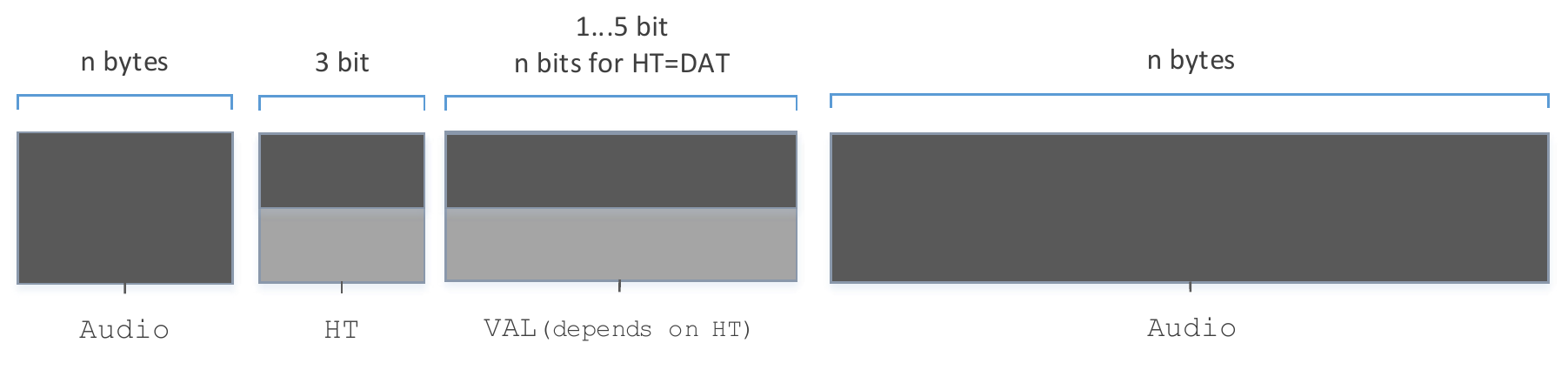, width=.95\textwidth}
\caption{Visualization of the general structure of a dynamic micro protocol header. Each packet contains only one header type (HT), i.e. no `pointer' to the start of next HT is required.}
\label{fig:DynHdrConcept}
\end{figure*}

All secondary header types are only included if a request is sent which results in smaller response and dummy messages. Requests are additionally shrinked in size by defining rarely used secondary header types as optional (indicated by the header type, see Tab.~\ref{tab:DynHdrFormat}). In comparison to the static header, DAT was replaced by a value field (VAL) of dynamic size that can indicate both, payload and the parameters for header fields, such as a length value for LEN or a version number for VER. The general structure of our dynamic header design is shown in Fig.~\ref{fig:DynHdrConcept}.

The start and the end of a transmission is signaled via request headers as \emph{Begin-of-Message (BOM)} and \emph{End-of-Message (EOM)} headers. The former is indicated by HT=REQ, VAL=0 and the latter is indicated by HT=REQ, VAL=1.

\begin{figure*}[t]
	\centering
	\epsfig{file=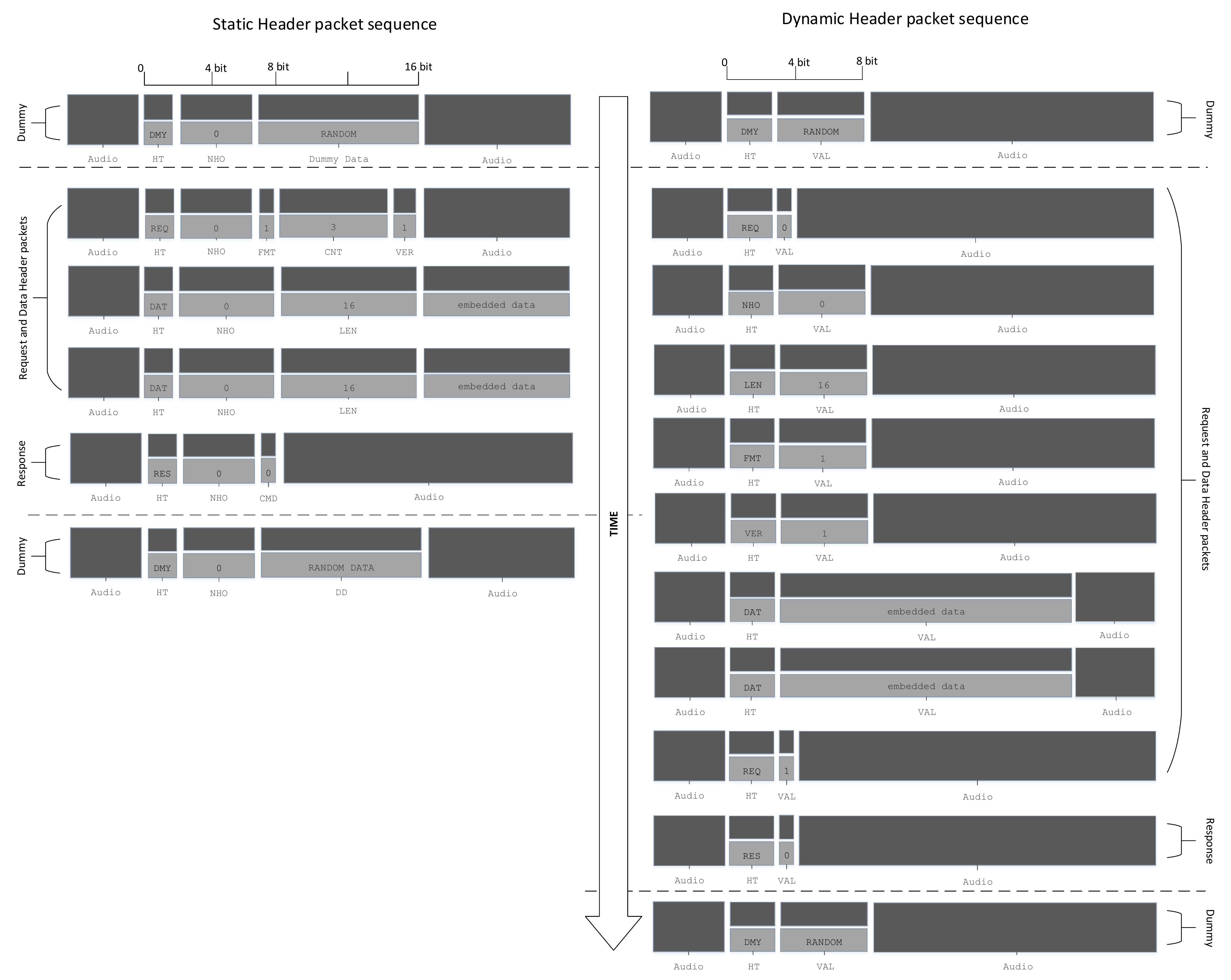, width=.95\textwidth}
	\caption{Comparison of static and dynamic micro protocol header designs.}
	\label{fig:CmpStatDynHdr}
\end{figure*}

We visually compare both design concepts, the static and the dynamic header, in Fig.~\ref{fig:CmpStatDynHdr}.

\section{Evaluation}\label{Sect:Evaluation}

We now evaluate the implementation of both the static and the dynamic micro protocol design. These results will serve as a basis for the design of our micro protocol engineering approach for unstructured carriers.

For evaluation purposes we chose to emulate an audio streaming service like VoIP by streaming a predetermined 5-minutes long .wav file. It contains audio recordings from the TIMIT \cite{TIMIT} continuous speech corpus which is one of the most widely used corpora in the speech recognition community. Both male and female voices speaking English were used. During the experiments, parts of the .wav input file were then inserted into the payloads of consecutive RTP packets and sent to the receiver where reassembling into a .wav file was performed. Then, the original and degraded files were compared with the use of PESQ (\emph{Perceptual Evaluation of Speech Quality}) \cite{ITUT} and the resultant MOS-LQO (\emph{Mean Opinion Squared-Listening Quality Objective}) was returned.

We compared the transmission of hidden data using both micro protocol designs with the same payload sizes for different variations of the LSB algorithm and different audio codecs. For both the static and the dynamic header design we performed transmissions with varying header structures and varying frequencies at which packets were sent. For this reason, we implemented three scenarios:

\begin{enumerate}
 \item A transmission of dummy packets with 16 bit payload.
 \item A transmission using a packet loop, each consisting of dummy packets, one REQ, few DAT and one RES packet. We sent requests  with 3x480 bit payload. As shown in Fig.~3, it was necessary to add additional packets to the dynamic header, namely for the header types NHO, LEN, FMT, and VER.
 \item A transmission using one REQ packet followed by as many DAT packets as feasible per request and header design.
\end{enumerate}

All requests were acknowledged with response packets in order to perform a bi-directional transmission. In detail each test scenario works as follows: First, the original PCM file will be read out. Second, the transmitter converts the file to the defined payload type and the transmitted bytes are recorded as raw bytes into a separate file. Third, the receiver reads the stream and decodes the embedded message if included. Fourth, the stream is saved in a .wav file. Fifth, streams with embedded messages and streams without embedded messages will be compared regarding changes at the bit level and their audio quality.

We compare the MSE (\emph{Mean Squared Error}), SNR (\emph{Signal-to-Noise Ratio}), PNSR (\emph{Peak Signal-to-Noise Ratio}), and MOS-LQO (\emph{Mean Opinion Score-Listening Quality Objective}) of our proof-of-concept implementation to evaluate the differences between audio transmissions without steganographic content and those with steganographic content.

\begin{figure*}[t]
\centering
\epsfig{file=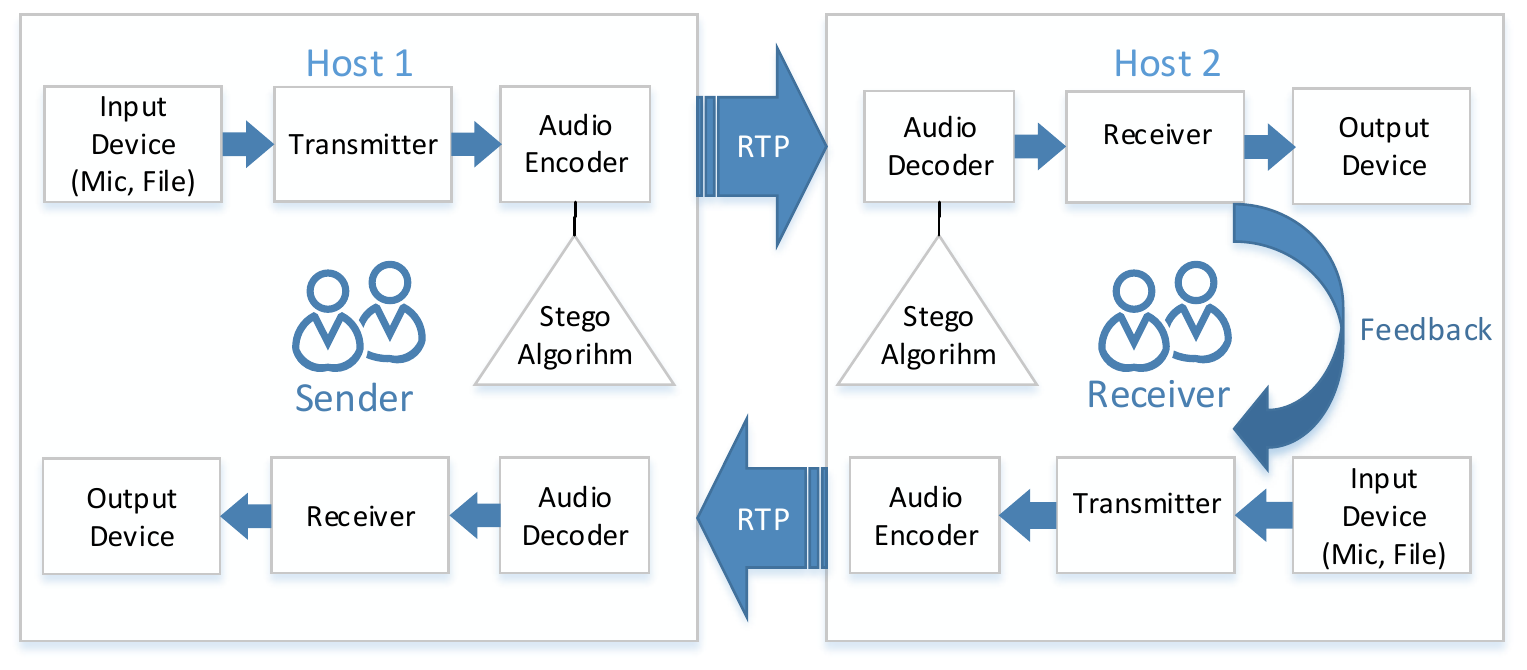, width=.95\textwidth}
\caption{Experimental setup for the evaluation of our micro protocol designs.}
\label{fig:ExpSetup}
\end{figure*}

Fig.~\ref{fig:ExpSetup} shows our experimental setup. Two hosts are communicating over a covert channel and the receiver provides feedback about the received steganographic data back to the sender. We applied the audio codecs ULAW and DVI; all transmissions are performed using the RTP protocol.

\subsection{Static Header Design}

\begin{figure*}[t]
\centering
\epsfig{file=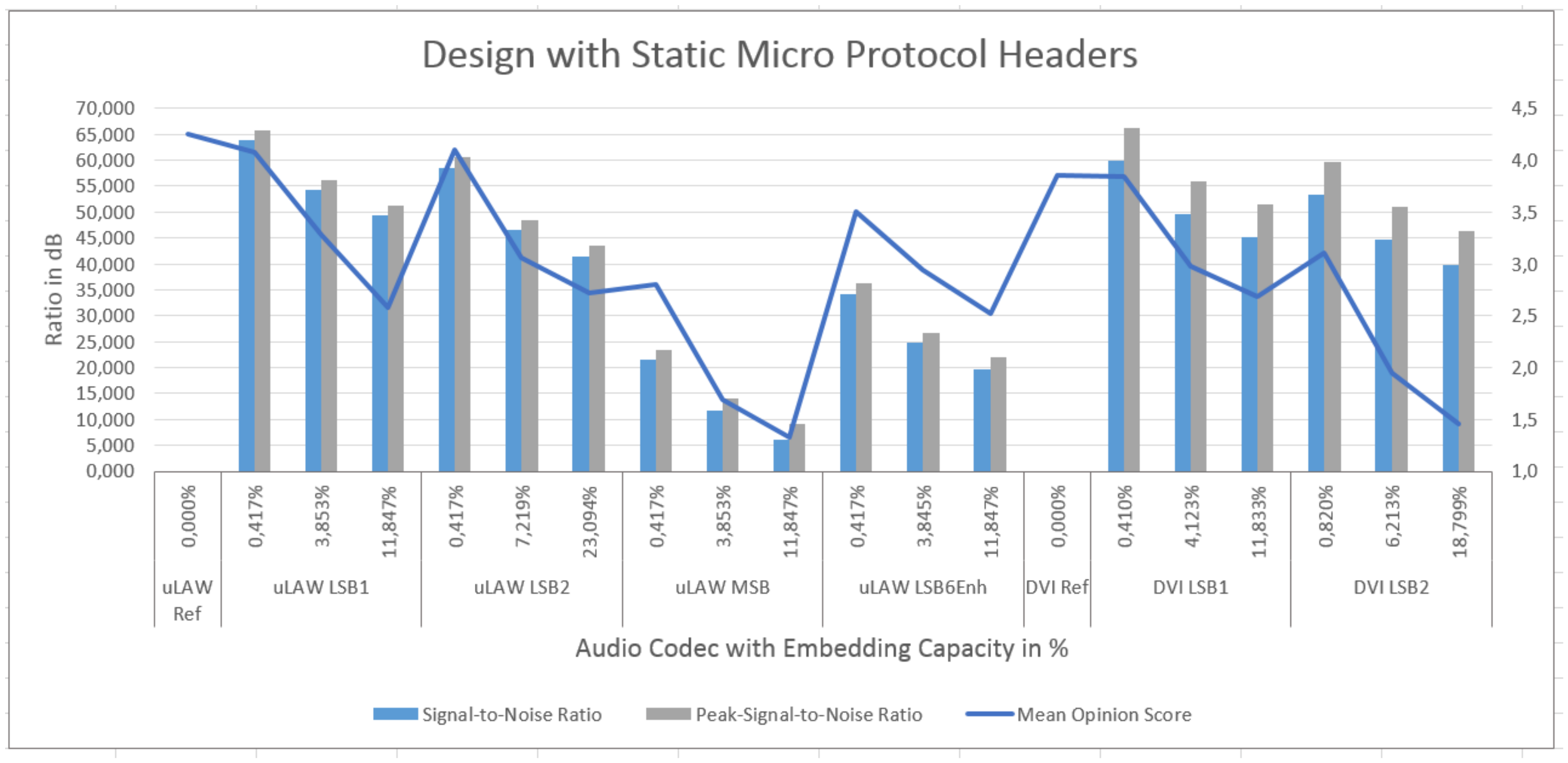, width=.95\textwidth}
\caption{Comparison of experimental results for the static micro protocol design.}
\label{fig:ResultsStatic}
\end{figure*}

The results for the evaluation of our static header design are shown in Tab.~\ref{tab:ResultsStatic}, which shows the MSE, SNR, PSNR and MOS-LQO values for different codecs and algorithms. Increasing 
the amount of embedded data results in a lower covertness, in a lower Mean Opinion Score and needs fewer packets with embedded payload per transaction. Tab.~\ref{tab:ResultsStatic} also reveals the insufficient performance for the MSB and LSB6Enh algorithms (the latter is utilizing the sixth least significant bit). We can conclude that changes of higher significant bits result, as expected, in a higher reduction of the voice quality when a static micro protocol is embedded. Fig.~\ref{fig:ResultsStatic} additionally visualizes the Signal-to-Noise Ratio, Peak Signal-to-Noise Ratio, and MOS values for the static micro protocol header case depending on the embedding algorithm used, codec and utilized embedding capacity.

\begin{table*} \centering
    \begin{tabular}{llccccc}
        \textbf{Codec} & \textbf{Algorithm}~ &  ~\textbf{Hidden Bits}  & \textbf{MSE}~ & \textbf{SNR (dB)} & \textbf{PSNR (dB)} & \textbf{MOS-LQO} \\
        \toprule
        \textbf{ULAW} & none (reference)	& 0\%	&	&	&	& 4.25	 \\
		(8Khz, 8bit, &	 LSB1 	& 0.417\%	& 0.017	& 63.881	& 65.861	& 4.080\\
	         mono)	& 		& 3.853\%	& 0.155	& 54,240	& 56.220	& 3.285\\
       		& 	& 11.847\%	& 0.478	& 48.357	& 51.339	& 2.577\\
        \cline{2-7}
        		& LSB2	& 0.417\%	& 0.057	& 58.554	& 60.543	& 4.099\\
        		&	& 7.219\%	& 0.918	& 46.509	& 48.505	& 3.054\\
        		&	& 23.094\%	& 2.952	& 41.398	& 43.429	& 2.720\\
        \cline{2-7}
        		& MSB	& 0.417\%	& 289.717	& 21.490	& 23.511	& 2.801\\
        		&	& 3.853\%	& 2581.209 & 11.714	& 14.013	& 1.687\\
        		&	& 11.847\%	& 7914.764 & 6.149	& 9.146		& 1.332\\
		\cline{2-7}
			& LSB6Enh & 0.417\%	& 15484 & 34.239	& 36.232	& 3.501\\
			&	& 3.845\%	& 135.814 & 24.713	& 26.801	& 2.938\\
        	&	& 11.847\%	& 412.661 & 19.661	& 21.975		& 2.523\\
		\hline
		\textbf{DVI} & none (reference)	& 0\%	&	&	&	& 3.85	\\
		(11Khz, 4bit, &	 LSB1 	& 0.410\%	& 0.016	& 59.927	& 66.226	& 3.847\\
         mono)	&	& 4.123\%	& 0.163	& 49.697	& 55.996	& 2.978\\
       		& 	& 11.833\%	& 0.473	& 45.082	& 51.382	& 2.678\\
        \cline{2-7}
        		& LSB2	& 0.820\%	& 0.071	& 53.318	& 59.617	& 3.101\\
        		&	& 6.213\%	& 0.517	& 44.686	& 50.995	& 1.951\\
        		&	& 18.799\%	& 1.548	& 39.899	& 46.232	& 1.457\\
		\bottomrule
    \end{tabular}
    \caption{Evaluation results for a static micro protocol header design.}
    \label{tab:ResultsStatic}
\end{table*}

\subsection{Dynamic Header Design}

The results for our experiments using the dynamic header are shown in Tab.~\ref{tab:ResultsDynamic}. Due to the additional header types (HT), the dynamic header consumes more space than the static header in case of small payload transfers. In other words, our highly fragmented header design produces an overhead of header data in comparison to the static header. For this reason, the dynamic header leaves slightly less space for the embedding of hidden payload\footnote{Backs et al.\ provide another dynamic header approach that provides slightly better results than a static header \cite{BacksEtAl}. Their header design provides only few fragments and thus decreases the overhead in comparison to a static header.} and the average embedding capacity for our test cases was higher in case of static headers (6.383\%) as for dynamic headers (5.188\%). However, if large payload is transferred, the overhead of the dynamic header is smaller than the overhead of the static header and thus provides better capacity. The MOS results are better in case of the dynamic header (in average 3.119 for dynamic header and 2.812 for the static header).

\begin{table*} \centering
    \begin{tabular}{llccccc}
        \textbf{Codec} & \textbf{Algorithm}~ &  ~\textbf{Hidden Bits}  & \textbf{MSE}~ & \textbf{SNR (dB)} & \textbf{PSNR (dB)} & \textbf{MOS-LQO} \\
        \toprule
        \textbf{ULAW} & none (reference)	& 0\%	&	&	&	& 4.25	\\
		(8Khz, 8bit, &	 LSB1 	& 0.208\%	& 0.008	& 66.896	& 68.875	& 4.242\\
        	 mono)	&		& 2.457\%	& 0.100	& 56,133	& 58.113	& 3.429\\
       		& 	& 11.549\%	& 0.469	& 49.437	& 51.420	& 2.636\\
        \cline{2-7}
        		& LSB2	& 0.208\%	& 0.025	& 62.234	& 64.214	& 4.248\\
        		&	& 4.704\%	& 0.602	& 48.348	& 50.338	& 3.290\\
        		&	& 22.785\%	& 2.915	& 41.454	& 43.484	& 2.722\\
        \cline{2-7}
        		& MSB	& 0.208\%	& 143.555	& 24.562	& 26.561	& 3.327\\
        		&	& 2.457\%	& 1674.369 & 13.712	& 15.892	& 1.946\\
        		&	& 11.549\%	& 7777.446 & 6.244	& 9.222		& 1.343\\
		 \cline{2-7}
        		& LSB6Enh	& 0.208\%	& 7.583		& 37.347	& 39.333	& 3.805\\
        		&	& 2.457\%	& 87.970 & 26.639	& 28.687	& 3.092\\
        		&	& 11.549\%	& 405.232 & 19.746	& 22.054	& 2.537\\
			
		\hline
		\textbf{DVI} & none (reference)	& 0\%	&	&	&	& 3.85	\\
		(11Khz, 4bit, &	 LSB1 	& 0.410\%	& 0.016	& 59.914	& 66.213	& 3.847\\
       		  mono)	& 	&	1.690\%	& 0.069	& 53.475	& 59.773	& 3.567\\
       		& 	& 6.918\%	& 0.280 & 47.362	& 53.661	& 3.302\\
        \cline{2-7}
        		& LSB2	& 0.410\%	& 0.029	& 57.264	& 63.562	& 3.439\\
        		&	& 2.526\%	& 0.209	& 48.627	& 54.930	& 2.885\\
        		&	& 11.093\%	& 0.918	& 42.185	& 48.504	& 2.494\\
		\bottomrule
    \end{tabular}
    \caption{Evaluation results for a dynamic micro protocol header design.}
    \label{tab:ResultsDynamic}
\end{table*}

\begin{figure*}[t]
\centering
\epsfig{file=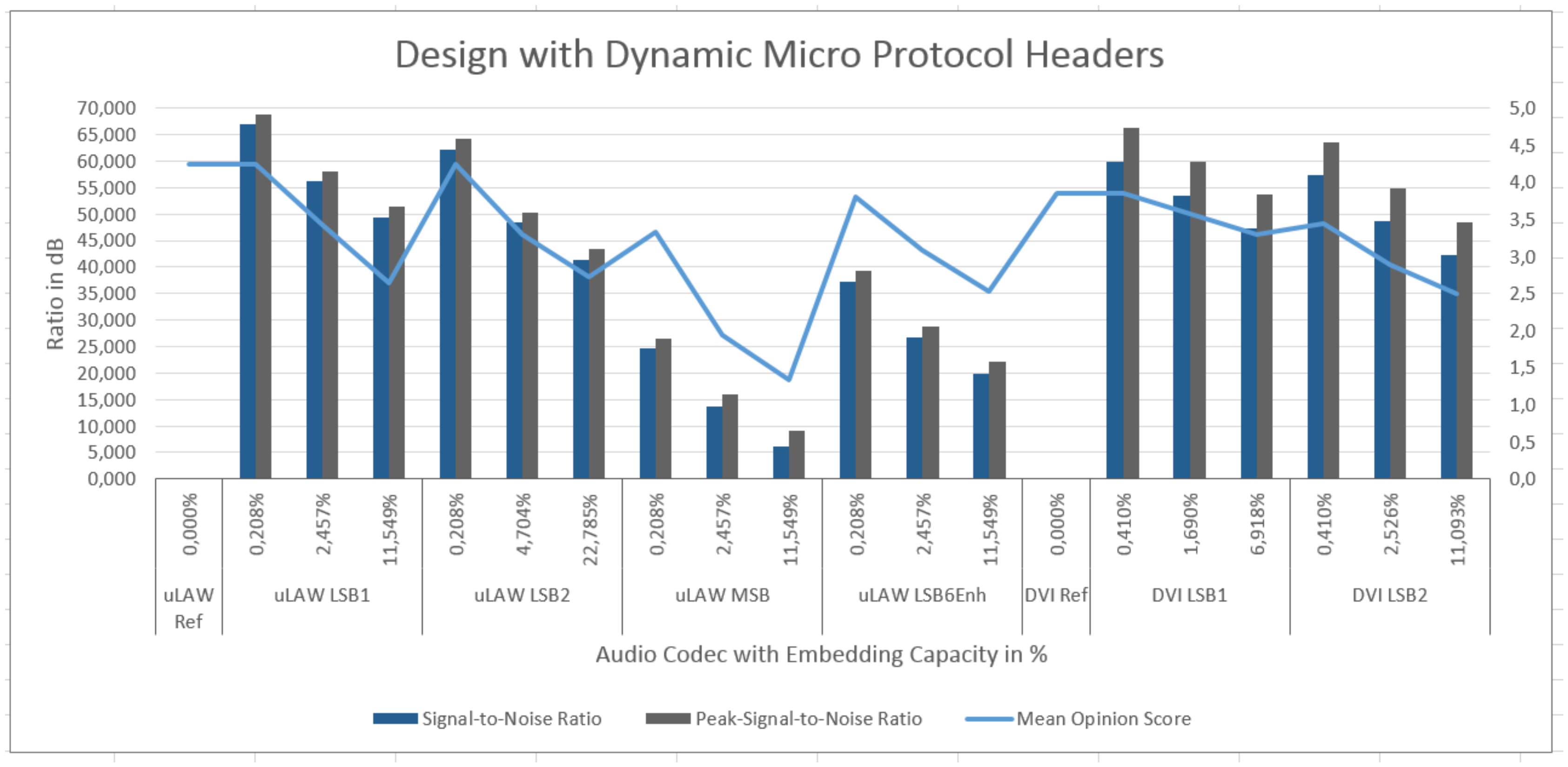, width=.95\textwidth}
\caption{Comparison of experimental results for the dynamic micro protocol design.}
\label{fig:ResultsDyn}
\end{figure*}

Fig.~\ref{fig:ResultsDyn} visualizes the Signal-to-Noise Ratio, Peak Signal-to-Noise Ratio and MOS values for the dynamic micro protocol header depending on the embedding algorithm, codec and utilized embedding capacity.

\section{Micro Protocol Engineering}\label{Sect:MPEngAppr}

Based on the results conducted in the previous experiments, we now derive recommendations for the design of micro protocols in unstructured carriers. We afterwards present our micro protocol engineering approach.

\subsection{General Recommendations}\label{Sect:MPEngAppr:GeneralRecomm}

Even small changes in audio transmissions can lead to the acoustic identification of quality reduction.\footnote{Regarding to Dickreiter et al., humans are able to recognize a voice inside a noised signal with a Signal-to-Noise Ratio of 6dB \cite{HandbuchTontechnik}.} At the same time, the embedding of larger amounts of data per packet without raising attention is difficult in highly compressed audio transmissions. A first recommendation is therefore that implementations of micro protocols should allow the selection of particular codecs on startup, depending on the amount of data to be transferred and other considerations such as time limits.

Our following recommendations are related to requirements, namely covertness, channel capacity, and reliability. Indeed, these requirements result in a multi-criteria optimization problem, as a higher channel capacity results in lower covertness, and may also influence reliability.

\paragraph*{Maximizing Covertness} An alternating offset for the initial `header type' field of the micro protocol header increases the hurdles for a warden to recognize patterns of steganographic transmissions and with it the hidden communication itself. We also recommend the introduction of dummy data packets (as forseen in our protocol designs). These dummy data are exchanged between hosts when no actual steganographic payload must be transferred. If no dummy data are transferred, the difference between actual steganographic transfers and audio transfers without steganographic content will be larger and thus may rise more suspicion. A small micro protocol header contributes additionally to the covertness and must be seen as a central element in any micro protocol engineering approach. We recommend the implementation of \emph{dynamic} headers as these result in a higher covertness.

\paragraph*{Maximizing Capacity} Codecs with lower compression allow the integration of more hidden data. Like in case of a maximized covertness, a small micro protocol header can contribute to a maximized capacity, too: the fewer space is required for embedding the header, the more space is left for embedding hidden payload. Advanced techniques for the embedding of audio steganography, such as Transcoding Steganography (TranSteg) \cite{Mazurczyk2014}, can be applied to increase the channel capacity. Moreover, a covert channel can utilize structured and non-structured elements of network packets simultaneously. For larger payloads, the covertness of the dynamic header provides better results and is recommended for this reason. For small payload transfers, the overhead of the dynamic header results in lower capacity than the static header approach. For this reason, the protocol engineer must decide -- based on the particular use case of the micro protocol -- which approach to chose if the capacity must be optimized.

\paragraph*{Reliability} In case of real-time streaming, the loss of audio packets occurs on a regular basis. If audio packets are lost, their steganographic content is lost as well. For this reason, we recommend the implementation of reliability mechanisms, such as ARQ (Automatic Repeat reQuest) or more advanced techniques. It is also imaginable to implement reliability not on a per-packet basis but instead on a per-$n$-packet basis, e.g. for every 10 packets. This heavily reduces the required number of acknowledgement messages.

Of course, knowledge about the reliability mechanism used can also be exploited to detect that the covert communication is taking place e.g.\ by intentionally introducing packet losses in order to observe and to correlate the responses. In such a case a more sophisticated reliability approach can be utilized as proposed e.g.\ by Hamdaqa and Tahvildari \cite{reLACK} which can be easily incorporated for audio steganography purposes. It provides a reliability and fault tolerance mechanism based on a modified $(k, n)$ threshold using Lagrange Interpolation and their results demonstrate that the complexity of steganalysis is increased.

\subsection{Micro Protocol Engineering Approach}

\begin{figure*}[t]
\centering
\epsfig{file=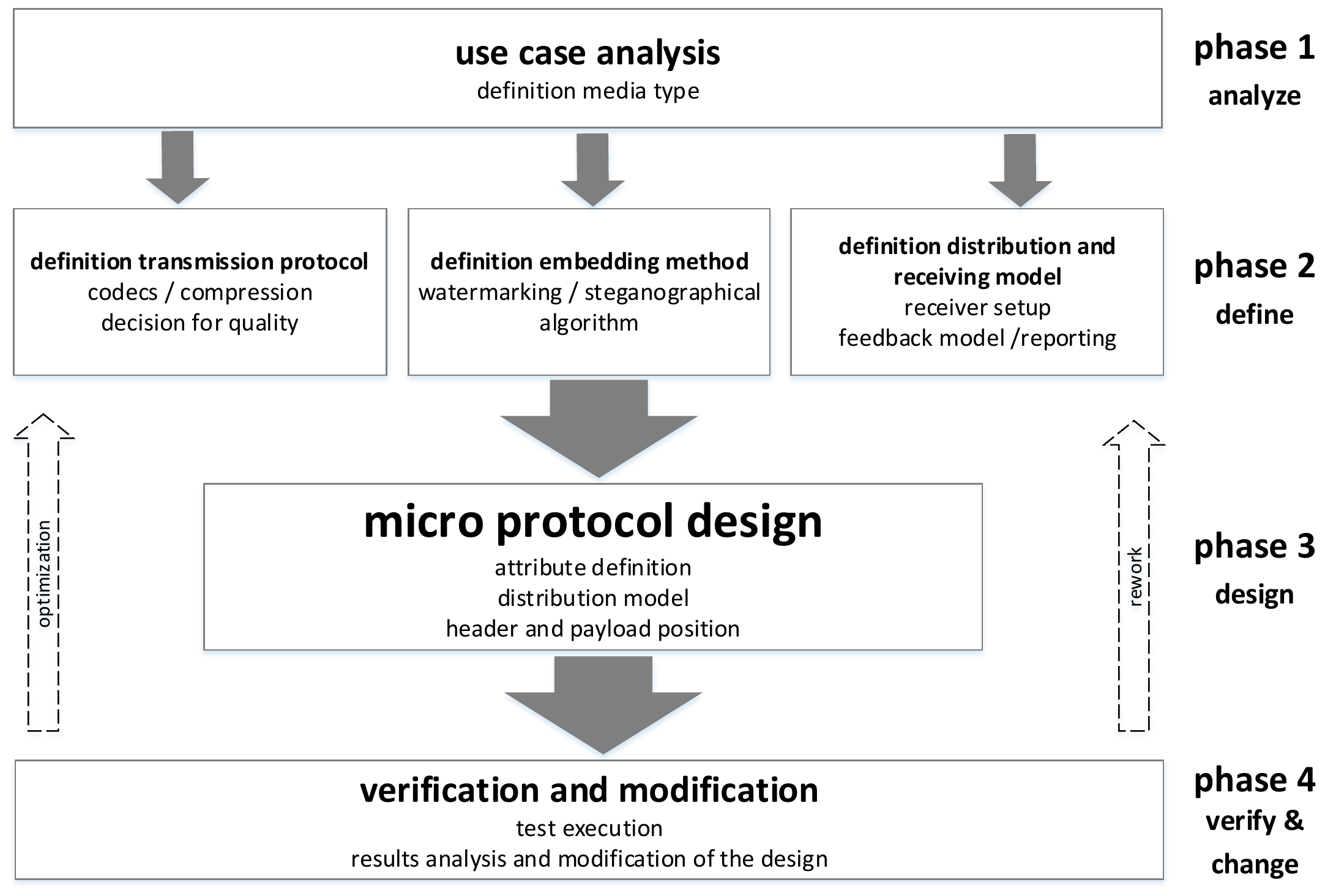, width=.95\textwidth}
\caption{Our phase model for the design of micro protocols in unstructured carriers.}
\label{fig:PhaseModel}
\end{figure*}

While Section~\ref{Sect:ReqAnal} discussed the requirements of our micro protocol implementation, we now present the requirements of our micro protocol engineering approach. The micro protocol engineering approach is based on a typical use case for which a micro protocol design will be created. The output of the micro protocol engineering process is the micro protocol design.

In \cite{CMS12}, Wendzel and Keller mention fundamental requirements of a micro protocol engineering approach, of which we consider the following two for our approach:
\begin{itemize}
 \item The micro protocol engineering approach must be applicable in practice.
 \item The most important criterion to fulfill by the micro protocol is to provide a high covertness. Therefore, the protocol engineering approach must support a maximized covertness of the resulting micro protocol.
\end{itemize}

We extend these aspects with the following additional requirements:
\begin{itemize}
 \item The micro protocol's transmission must be reliable, i.e. possess a feedback channel or another means for error correction.
 \item The micro protocol should be adaptable to different codecs used for the unstructured carrier.
\end{itemize}

Based on these requirements, we derive a four-phase model for the micro protocol designing process. We limit the required workload of the protocol engineer and design the particular steps in a practically applicable manner. The model is shown in Fig.~\ref{fig:PhaseModel}. A switch between the phases is feasible at any time to optimize the micro protocol design.

\paragraph*{First Phase} In the first phase, a use case analysis must be performed. The use case analysis defines the requirements (e.g.\ high channel capacity) which are linked to that use case. For instance, some micro protocols must be designed for high-throughput transmissions while others may only need to transfer a few bits per second. A sample use case could be the transfer of illicit video data over slow or non-reliable connections. Moreover, it must be defined whether a real-time transmission or the transmission of data on-demand will be required and which transmission media will be used. Finally, it must be defined whether the steganographic transmission must be reliable.

\paragraph*{Second Phase} Based on the requirements definition in the first phase, the second phase defines the network protocols, the media codecs, the compression algorithm used and the steganographic embedding algorithms to be used. For instance, if the requirement is a real-time transmission, the RTP protocol can serve as a network protocol.

If steganographic data is embedded into the audio data \emph{after} the compression took place, the hidden data may not be subjected to change during transmission. In such a case \emph{spatial-domain} methods such as Least Significant Bits (LSB) or adding information to silence sections of the voice transmission can be applied. If changes of the steganographic content can be performed within the transmission, it may be more suitable to select \emph{frequency-based} methods such as Phase Coding, Spread Spectrum, or Echo Data Hiding.

Finally, if reliability was defined as a requirement within the first phase, a means must be defined to realize reliability, e.g. by adding sequence numbers and a TCP-like reliability algorithm or more covert approaches like described in Section~\ref{Sect:MPEngAppr:GeneralRecomm}.

\paragraph*{Third Phase} First, the protocol engineer must determine the available space to place hidden data into the unstructured carrier. The available space is a result of the audio codec used. The resulting value is used as a limit for the size of the micro protocol header.

Secondly, the requirements for covertness and channel capacity must be addressed. As discussed in Sect.~\ref{Sect:MPEngAppr:GeneralRecomm}, dynamic header designs can increase the covertness and -- for larger payload -- the capacity of a transmission while for small payload transactions, a static header will provide a better capacity. After this decission is made, the required size for the micro protocol header must be determined. All necessary features, such as the capability to transfer dummy data or to provide reliability, must be integrated into the header (represented by single or multiple bits as header fields). If the micro protocol consumes too much space, features must either be reduced or the micro protocol must be designed in a more dynamic way. If both actions are not feasible, a carrier that provides more space must be selected, which leads back to the second phase.

Thirdly, the protocol engineer subtracts the micro protocol header size from the available hidden storage space within the carrier. The result is the available space for the hidden payload per packet. If the available space is insufficient to transfer the necessary number of bits per second in the given use case, the engineer must re-design the micro protocol or must change definitions within the second phase.

Fourthly, the protocol engineer must specifiy how the start of micro protocol data within the carrier is indicated. This can be done by defining a fixed offset for the first packet and defining sequences to identify data.\footnote{The larger the allowed next header offset values (header type NHO), the less space will be available per packet (see third step).}

Fifthly, the micro protocol must be implemented in a proof of concept system.

\paragraph*{Fourth Phase} Finally, the micro protocol design must be verified. For this reason, a proof of concept code must be implemented and tested. Passive wardens, either as humans or as software, should be used to assess the non-detectability of the covert communication. Moreover, it must be evaluated whether the embedding of the steganographic data can result in behavior that is not conforming to a codec's specification. This last step includes ensuring that attributes defined in the audio codec (e.g., the packet capacity, specified sampling rate, bit depth or needed audio transformation steps) are not violated by the embedded micro protocol. A simple method is to verify the correct processing of the modified streams by a common player software, e.g. a streaming client.

\section{Conclusions and Future Work}\label{Sect:Concl}

We demonstrated the embedding of a micro protocol into an unstructured carrier, namely audio streaming over the network. The micro protocol was implemented in a static and a dynamic way to compare the resulting performance and impact of both designs on the audio signal. Based on this feasibility study and its evaluation, we developed a micro protocol engineering approach for unstructured carriers. Our micro protocol engineering approach extends the existing approaches which were solely focusing on structured carriers.

Micro protocols can be designed in a different way as proposed in this article, e.g. more header functionality can be integrated and alternative audio codecs can be applied. However, our discussed designs are highly dynamic and, for this reason, represent a generic design approach that can easily be modified by introducing new or replacing present header types.

At the moment, our framework does not address the micro protocol's temporal behavior. However, the micro protocol's temporal behavior should at least be adjusted closely to the temporal behavior of the carrier (e.g.\ inter packet gaps) --- otherwise its temporal behavior may lead to detectability \cite{KaurEtAl:MPCountermeasures}. The micro protocol engineering framework will thus be extended for this purpose.

\bibliographystyle{abbrv}
\bibliography{mp_unstr_car}

\end{document}